\documentclass[aps,prb,preprint,amsmath,amssymb,groupedaddress, showpacs]{revtex4-1}

\usepackage{graphicx}
\usepackage{bm}

\begin{document}

\title{A Deformable Model for Magnetic Vortex Pinning}

\author{J.A.J. Burgess$^{1,2}$, J.E. Losby$^{1,2}$, and M.R. Freeman$^{1,2}$}
\affiliation{$^{1}$ Department of Physics, University of Alberta, Edmonton, Alberta, Canada T6G 2E1}
\affiliation{$^{2}$ National Institute for Nanotechnology, Edmonton, Alberta, Canada T6G 2M9}

\date{\today}

\begin{abstract}
A two-parameter analytical model of the magnetic vortex in a thin disk of soft magnetic material is constructed. The model is capable of describing the change in evolution of net vortex state magnetization and of core position when the vortex core interacts with a magnetic pinning site. The model employs a piecewise, physically continuous, magnetization distribution obtained by the merger of two extensively used one-parameter analytical models of the vortex state in a disk. Through comparison to numerical simulations of ideal disks with and without pinning sites, the model is found to accurately predict the magnetization, vortex position, hysteretic transitions, and 2-D displacement of the vortex in the presence of pinning sites. The model will be applicable to the quantitative determination of vortex pinning energies from measurements of magnetization.
\end{abstract}


\maketitle

\section{Introduction}

Interest in magnetic vortices\cite{CowburnPRL1999, ScienceMFM} in thin disks has grown dramatically over the past two decades, as these are fundamental physical systems with direct applications to technology \cite{RaceTrack}. Topological structures such as vortices are stable, manipulable objects that show promise as logic elements or storage media in spintronics applications. The thin soft magnetic disk, the prototypical system containing a vortex, has therefore become an extensively investigated system. Properties studied include structure\cite{UsovPeschany94, ScienceSPSTM},  dynamical modes\cite{Slon1984, Crowell2003, Acremann2004}, annihilation\cite{NovoCA2001, Burgess2010}, and creation\cite{NovoCA2001, DavisNJP, GMihaj, GusSlowDynamics}. As each aspect of vortex physics is probed experimentally, and considered for technological applications, theoretical understanding via simulation and modeling is also advanced. Modeling is particularly important in the case of the thin ferromagnetic disk as it presents a well-defined system amenable to description by an analytical approach. Here a two parameter analytical model is developed to enable qualitative and quantitative computation of vortex pinning effects in disks. 

The interaction of vortex cores or domain walls with film inhomogeneities has been a topic of significant recent interest. Geometric defects or magnetic impurities can increase or decrease the energetic cost of the topological structure\cite{Toscano2011}, creating preferential locations or altering the magnetization distribution.  In the disk system, direct observations of vortex state pinning have been made with Lorentz microscopy \cite{Uhlig2005} while the effect on vortex gyration has been observed with time-resolved magneto-optical Kerr effect microscopy\cite{Crowell2006, CrowellPRB2010, CrowellPRB2012, Crowell2012aXriv} and electronically\cite{Klaui2010}.  Incorporation of pinning potentials into existing analytical models has permitted a qualitative description of the position of the vortex and its reduced displacement susceptibility\cite{RVMpin}.  This approach is insufficient for quantitative applications. Recent work using nanomechanical torque magnetometry has provided direct observation of the Barkhausen steps associated with jumps in core position\cite{BurgessFraser}, necessitating the development of a model that permits a quantitative description of pinning effects.  Physically one expects two clear contributions to the magnetic susceptibility of a pinned vortex, one from translation of the entire magnetization distribution and one from deformation. A two-parameter analytical model that uses a dipole-exchange spring picture of a vortex to capture both contributions is presented here, and demonstrated to reproduce numerically simulated quasistatic vortex pinning behavior.

\section{The Vortex State and Existing Models}

In zero field, the vortex state in a disk represents the ground state configuration for a wide variety of disk aspect ratios. Over most of the disk, a circularly symmetric in plane magnetization distribution maintains magnetization tangential to the disk boundary and reduces dipolar energy. This necessitates a higher exchange energy relative to the uniformly magnetized state and results in an out of plane magnetized core at the disk center with high energy density. 

The vortex state ansatz was first developed for the magnetic disk by Aharoni in 1990\cite{Aharoni}. Further work by Usov and Peschanny\cite{UsovPeschany94} determined an exchange optimized functional form of the core magnetization profile.  Good agreement between this model and simulation\cite{Scholz2003},  as well as experimental observation\cite{ScienceSPSTM}, was found.  This work considered the vortex ground state, at zero field. Computation of the evolution of the state with field presents a more challenging problem. 

As field is increased the segment of the disk with magnetization aligned with the field grows, causing other features of the circularly symmetric distribution to shift in response. The expansion can be occur in two ways: the entire magnetization distribution may translate orthogonal to the field to create a larger section aligned with the field, or, the region favored by the field direction may simply expand within the disk disrupting the circular symmetry of the magnetization distribution about the core position. Both have energetic costs and will contribute to the increasing magnetization of the disk. 

\begin{figure}[H]
\includegraphics[scale=0.8]{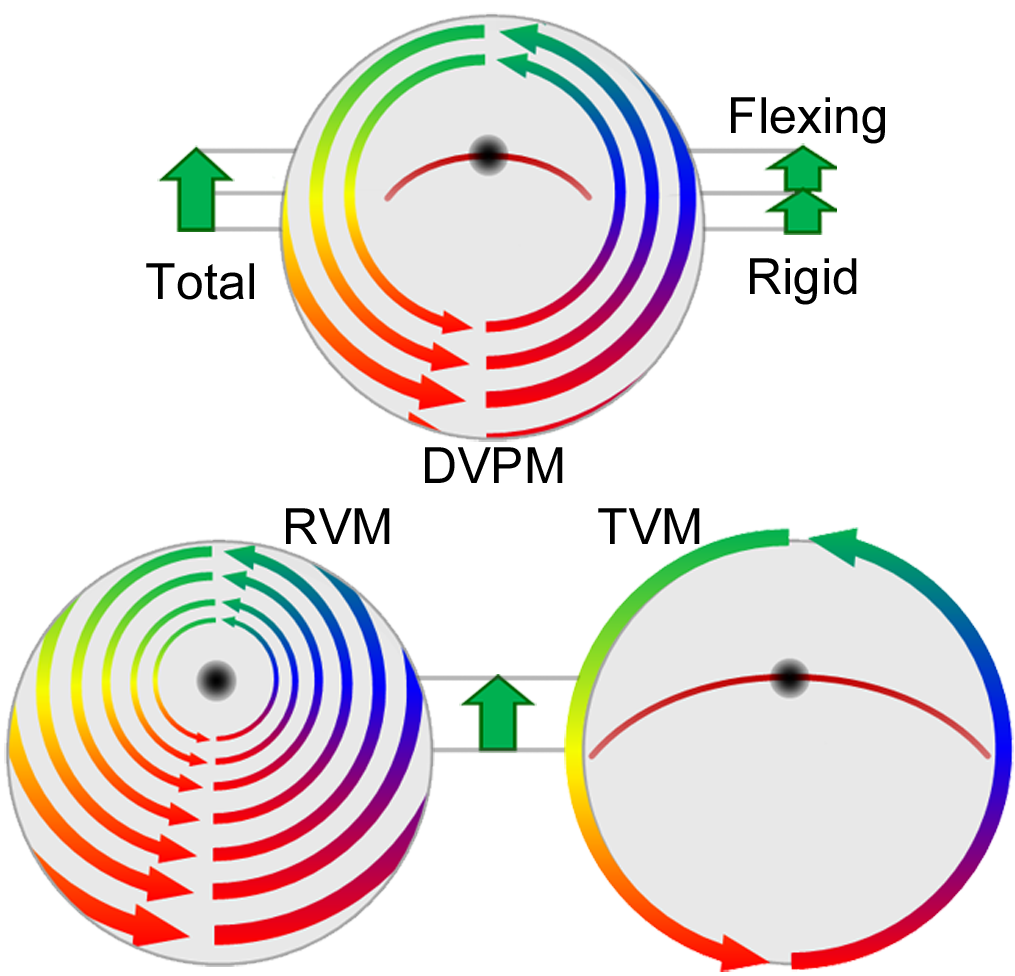}
\caption{\label{fig1} (Color available in online version.) Schematics depicting the evolution of the magnetization distribution as the vortex core is displaced. The arrows within the disks, as well as the color scale, indicate the in-plane direction of the magnetization. At top is the DVPM which has two regions: an outer shell described by the RVM and an inner shell described by the TVM. Both components make contributions to the magnetization and vortex displacement allowing the DVPM to capture more complex behavior. By contrast the RVM (lower left) employs a parameter that governs both vortex and magnetization distribution displacement, while the TVM (lower right) has a single parameter that determines both magnetization distribution flexing and vortex displacement. }

\end{figure}

The translation has been addressed with the development of the Rigid Vortex Model (RVM) which is considered to originate with Usov and Peschanny's work.  Subsequent work \cite{GusPRB, GusNovo} considered the displacement of the core with field, and annihilation field, using a model that rigidly translated the magnetization distribution developed by Usov and Peschany (Figure \ref{fig1}). This is known as the Rigid Vortex Model (RVM). Recent extensions of the RVM include higher order versions developed to describe the susceptibility of the displaced vortex analytically\cite{Burgess2010}. Concurrently, models approximating the deformation of the magnetization distribution were developed\cite{TVMJMMM, GusMetPRB}. This class of model is equivalent to the influence on the magnetization distribution of a second vortex moving from infinity to the edge of the disk, and is called the Two Vortex Model (TVM).  Versions of this model have been applied to calculating the stability of the vortex state in a disk with moderate success\cite{TVMJMMM, GusMetPRB} and, with greater success, to predict frequencies of dynamic modes\cite{TVMJAP, BuchananTVM2007, GusTVM4thPRB2010}. The most successful version of this model is one that maintains a perfect tangential boundary condition, prohibiting any translation of the magnetization distribution and considering only deformation.

The complete physical picture of the vortex state under an applied field involves translation and a continuous deformation of the magnetization distribution from the core all the way out to the edge of the disk. The deformation displaces the core, as is computed in the TVM, but it also decreases the energetic cost of translating the magnetization distribution rigidly by partially maintaining the tangential boundary condition. The bulk of the deformation can be considered as a widening of the section of the magnetization distribution aligned with the field. However the maintenance of the tangential boundary condition means that this widening is more prominent near the center of the disk. This leads to curvature in the deformation, and also to core deflection. In the absence of pinning, translation and distortion both increase monotonically under the influence of field. The entire distribution shifts, but the distribution is not rigid, and the core displaces an additional amount, ahead of the translation. 

The presence of a pinning site interaction with the vortex core means that the core position is no longer dictated solely by torques exerted on the core by the surrounding in-plane magnetization distribution and that there are preferential locations for the core in the disk. When the core is in a preferential site, the energy cost of further core displacement is increased and this necessarily influences the nature of the deformation. Under increasing field, the magnetization of the disk will continue to grow despite the pinning. This results in the magnetic moment away from the core increasing preferentially, favoring the translational mode of displacement because it results in a larger magnetic moment near the edge of the disk. Simultaneously, the displacement of the core due to distortion decreases (relative to the translation), maintaining the core in the preferential location. This allows some of the circular symmetry of the initial magnetization distribution to be restored. 

\begin{figure}[H]
\includegraphics[scale=0.8]{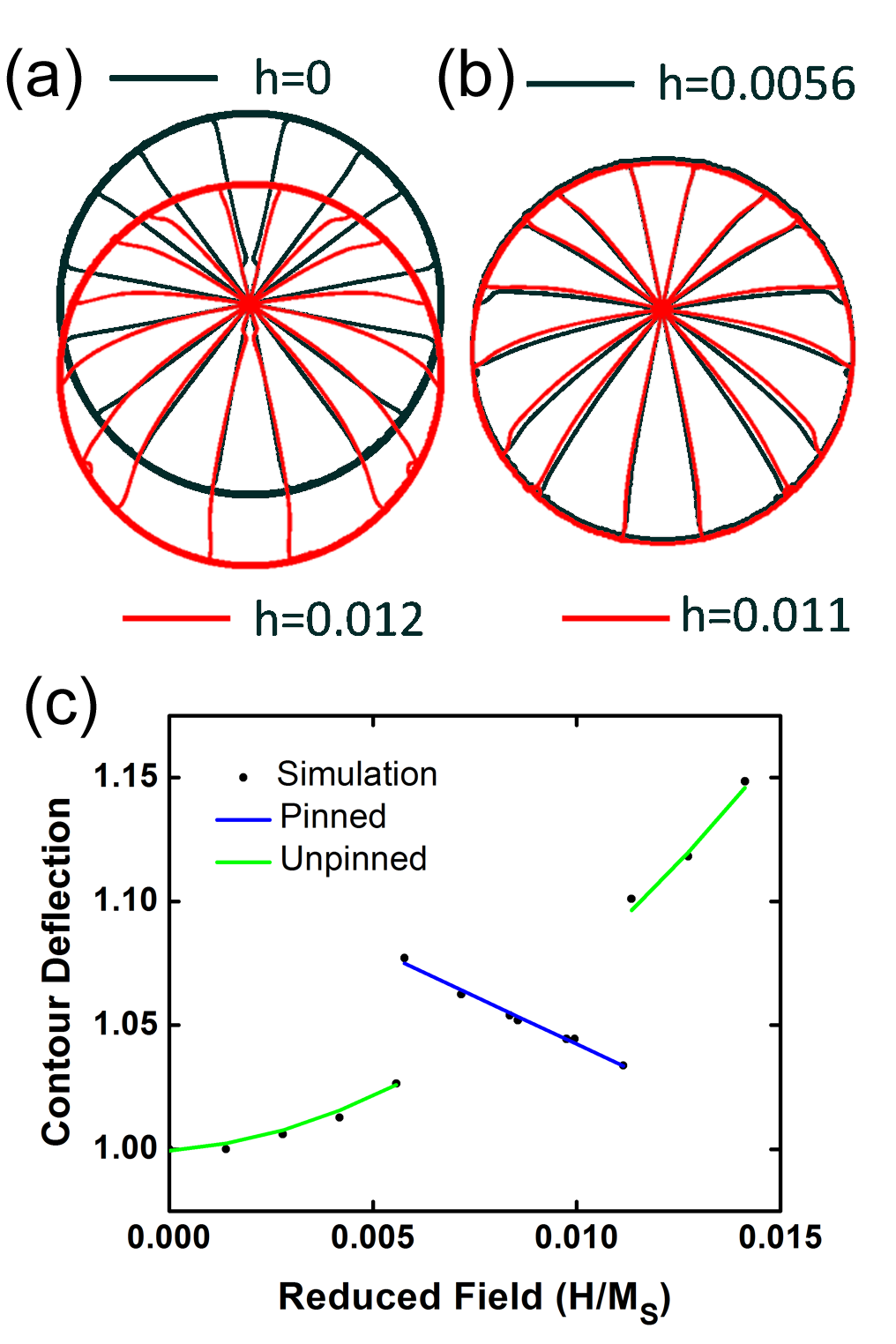}
\caption{\label{figf} (Color available in online version.) (a) Contours of constant $|M_y|$ value are computed from a simulation for two different fields and overlaid such that the vortex core positions overlap.  (b) The same type of plot is produced for the core trapped in a pinning site. The pinned contours qualitatively shift the opposite direction between low field and high field compared to case (a). (c) The normalized and averaged angular deflection of contours is shown as a function of field as the vortex moves through the pinning site. The angles are measured between the vertical ($|M_x|=1$) line and the intersection of the contours with a circle of radius $R/2$ centered on the vortex core. This reflects the net deformation of the magnetization distribution inside of $R/2$. For a pinned core and increasing field, the angular deflections reduce, partially restoring the circular symmetry of the zero field vortex state. In addition to the gentle bending of the contour lines throughout the disk, there is additional, sharper bending near the edge of the disk. This bending directly relates to maintenance of the tangential boundary condition and the reduction of the energetic cost of translation.}

\end{figure}

The observation of the restoration of circular symmetry in micromagnetic simulations incorporating pinning verifies the applicability of the combined translation and deformation picture. The deformation can be visualized by considering contours of constant magnetization, which bend away from radial lines as the magnetization deforms from the circularly symmetric vortex state. The widening of the section magnetized parallel to the field manifests as shifts in the angles of the contour lines, and an introduction of curvature. Near the center of the disk, the contours deflect only gently, however this gentle deflection extends all the way out to the edge of the disk, showing a significant deformation-based core displacement. Near the edge of the disk more extreme bending occurs, this being directly associated with the maintenance of the boundary condition and is significant in the reduction of the energetic cost of translation. Examining the gentle bending near the center, the contours exhibit a deflection with a non-monotonic evolution as the vortex traverses the pinning site (Figure \ref{figf}). During pinning, the contours begin to return to the original (zero field) angles of deflection, partially restoring the circular symmetry of the magnetization distribution. The more extreme deformation near the disk edge continues to grow, indicating translation continues during this time. Simulation movies presenting the evolution of the magnetization distribution in the unpinned and pinned states are available in the supplementary material (Movies M1 and M2).

The change in deformation represents a significant energetic influence on the core. The flexible nature of the magnetization distribution acts as a combined dipole-exchange spring. The spring can absorb energy, allowing a core to jump ahead to a preferential site. Similarly, stored energy lowers as the core is trapped in the pinning site, permitting the core to stay in the site longer. This has a significant effect on observed hysteresis in pinning sites, as well as on computation of depinning energy barriers. Both the RVM and TVM link the computed model magnetization directly to the vortex core displacement in the disk using only a single parameter\footnote{The TVM uses a second parameter to describe distortions of the core profile under deflection, however this parameter does not have a significant influence on the overall disk magnetization or core displacement. It is therefore unsuitable for describing the influence of the skirt magnetization on the core.}. This limitation renders the models incapable of computing, or even qualitatively describing the non-monotonic evolution of the deformation. However a combination of these two models may be constructed, modeled after the observed magnetization distribution and incorporating both rigid translation and flexible deformation.

\section{The Deformable Vortex Pinning Model}

The full solution can be pictured as a flexible distribution (TVM-like) that also has translation (RVM-like). To solve the complete model of a realistic disk presents a significant challenge: both the exchange and demagnetization energies yield results that convolve deformation and displacement. Furthermore, the character of the deformation will be influenced by the amount of translation. Therefore the full solution is fundamentally different than the exchange energy minimizing TVM-based models. One approach to simplify this problem is to develop a model that mimics the physical situation, while also providing a method of decoupling the effects of translation and deformation. 

This is the approach followed here by use of a piecewise model that we call the Deformable Vortex Pinning Model (DVPM). The concept is to have a region describing the dominant magnetic moment that develops in the outer section of the disk, coupled by a flexible region to the core, and where both the outer magnetic moment and the flexible region remain fully described analytically. To do so, the disk is divided into two circularly symmetric regions. An outer annular section described by the RVM  surrounds an inner section described by the TVM. The outer section provides rigid translation. The inner section translates with the outer annulus while providing deformation and consequently, a parameterization of the dipole-exchange spring. To construct the model in detail, we first review the component models. For each model the total energy is composed of the exchange, demagnetization, and external field energies as summarized below.  Magnetocrystalline anisotropy is neglected. 

\subsection{The Rigid Vortex Model}

The RVM is derived by considering the zero field vortex magnetization distribution\cite{Aharoni, UsovPeschany94} to be immutable, and then translating that distribution relative to the physical boundary of the disk (Figure \ref{fig1}). The normalized total energy for the 3rd order RVM \cite{Burgess2010} of a disk with a radius R and thickness L as a function of the reduced field $h=H/M_S$ is given as 

\begin{equation}
\frac{E_{tot}}{\mu_0M_s^2V}= \frac{\beta}{2}b^2-h (b-\frac{b^3}{8}),
\label{eq:Ervm}
\end{equation}

where the normalized core displacement $s =\Delta r/R$ is equal to $b$, $\beta = F_c(L,R) - R_o^2/R^2$ is a constant describing the demagnetization energy and exchange energy with $F_c(L,R)$ representing the susceptibility-corrected demagnetization factor computed for the uniformly magnetized disk\cite{DXChen}, and $R_o = \sqrt{2A/\mu_o M_S^2}$ is the exchange length.  The energy is normalized by $\mu_o M_S^2 V$ where V is the disk volume. 

The incorporation of the susceptibility correction to the demagnetization factor, the so called $\mu^*$ correction\cite{Kittel1949,Chickazumi}, is critical to the success of the model in application to a disk composed of a soft magnetic material (Figure \ref{fig2} a, inset), but is often neglected.  The susceptibility correction was introduced to account for the overestimate of the demagnetization factor computed for a uniformly magnetized particle assuming a rigid magnetization. In reality, the magnetization in the particle will deform away from the uniform state, paying a small energy price from introducing volume demagnetization charges but causing a net energetic reduction by decreasing the edge demagnetization energy (see reference \cite{Chickazumi} p. 437 Figure 17.6). It is this flexing that the correction takes into account. In the RVM, the demagnetization factor used is equal to that of a uniformly magnetized disk, and consequently it follows that introducing the susceptibility correction will parameterize the reduction in demagnetization energy enabled by small deformations of the magnetization distribution away from the rigid circular symmetry of the vortex. 

In an ideal (permalloy-like) material, the susceptibility is considered as infinite. Typically low coercivity ferromagnets can be treated as having infinite susceptibility ($>$100). In this case, we can estimate the maximum possible error in the demagnetization factor resulting from this approximation as $<0.8\%$ by considering the disk to have effective susceptibility equal to that of iron ($\sim49$)\cite{Kittel1949}. With an estimated susceptibility, the corrected demagnetization factor may be computed \cite{DXChen}, meaning that the correction gives back a fixed demagnetization factor for a known disk size. Therefore, the correction does not introduce a additional fit parameter. The logic of applying the correction to a rigid vortex distribution is borne out by the significant improvement in the performance of the RVM in predicting the vortex state magnetization and the core position (Figure \ref{fig2}). The improvement in the performance of the RVM stemming from the susceptibility correction makes it clear that the correction parameterizes immediate bending of the magnetization near the edge of the disk, as well as deflection through the volume of the disk. This is what allows the corrected RVM to provide a good estimate of both net magnetization and total vortex core displacement despite treating only translation. 

Solving the model permits calculation of the vortex displacement as a function of field, $b_o(h)=(-4\beta + 2 \sqrt{4\beta^2+6h^2})/3h$, and the magnetization $m_o(h) = M_o(h)/M_S=b_o-b_o^3/8$.  Removing the third order term in the magnetization, and consequently the external field energy contributing to $b_o(h)$, reduces the model to the second order RVM\cite{GusPRB, GusNovo} with $m_o(h) = b_o(h)  = h/\beta$.  

\subsection{The Two Vortex Model}

The TVM is derived by setting a boundary condition and computing the magnetization distribution that minimizes the exchange energy for a given vortex core displacement\cite{TVMJMMM, MetlovPRL}. Here the no side charges version of the model\cite{TVMJMMM,TVMJAP} is applied and the magnetization is held tangential to the disk edge.  Contributions from the core are neglected here, which is a reasonable approximation for disks with a radius significantly larger than the core radius\cite{TVMJMMM}. In the same form as used for the RVM, the total normalized energy may be written down for the TVM,  

\begin{equation}
\frac{E_{tot}}{\mu_0M_s^2V}= \frac{\alpha}{2}a^2-\xi ha,
\label{eq:Etvm}
\end{equation}

where the normalized core displacement $s=\Delta r/R$ is equal to $a/2$, $\alpha =R F_1(L,R)/L - R_o^2/2R^2$ incorporates the demagnetization energy and exchange energy, and $\xi$ is a constant ($\sim 10/29$). The function $F_1(L,R)$ is an equivalent demagnetization factor describing the volume magnetostatic charges resulting from flexing of the magnetization distribution and is approximated as $k (L/R)^2$ with $k=0.08827$. As before, minimization with respect to $a$ allows computation of $a_o(h)=\xi h/\alpha$ and magnetization $m_o(h)=\xi a_o(h)$. 

\begin{figure*}[H]
\includegraphics[scale=1.0]{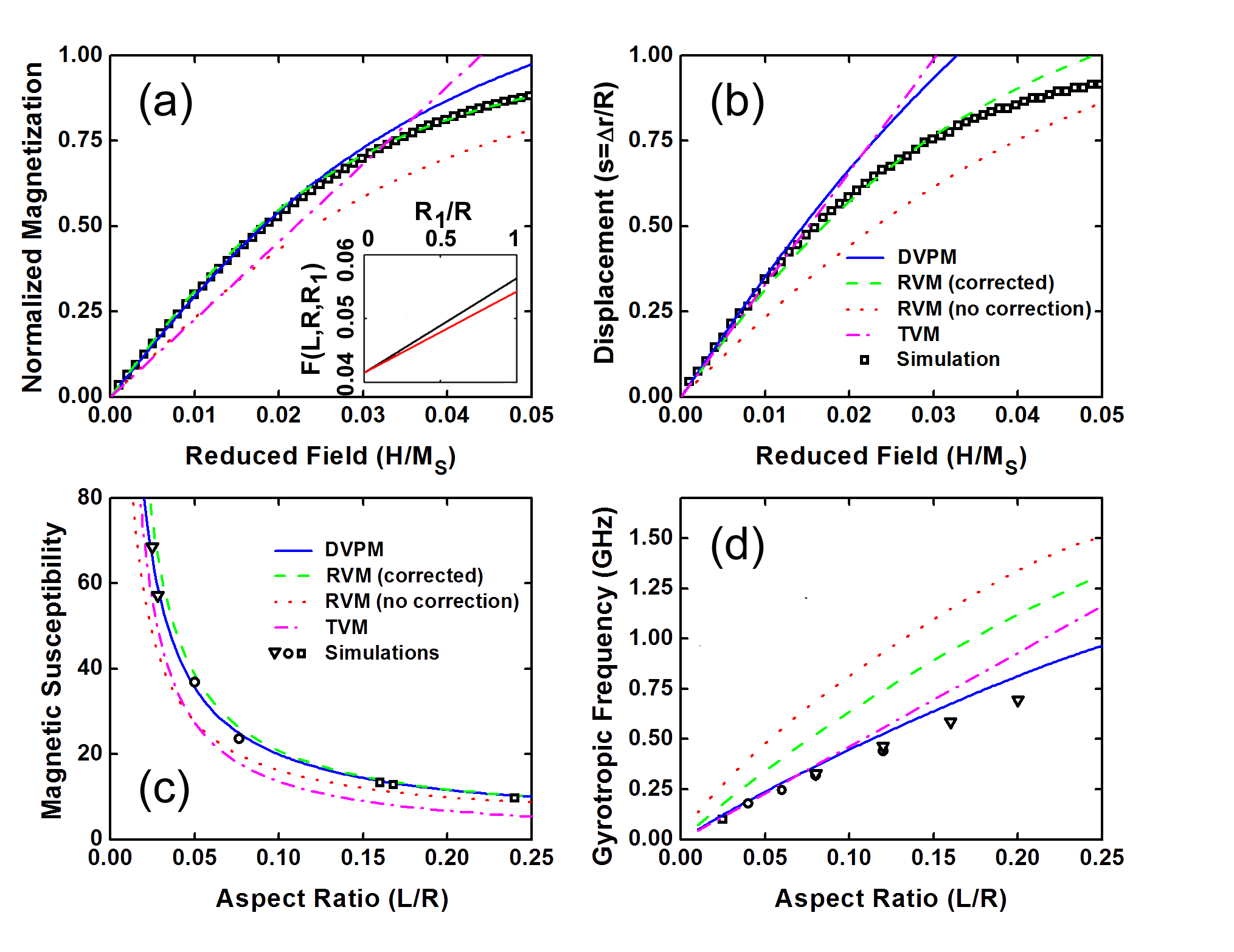}
\caption{\label{fig2} (Color available in online version.) (a) The computed m-h curves from four models (RVM corrected and uncorrected, DVPM, TVM) are compared against a simulation of a 1$\thinspace \mu$m diameter, 30$\thinspace$nm thick disk with $M_S$=800$\thinspace$kA/m. (b) The computed normalized vortex displacement as a function of field is compared against the simulation. The legend inset in panel (b) applies to both (a) and (b). Only the RVM with a susceptibility-corrected demagnetization factor and the DVPM describe both position and magnetization accurately for displacement $s<1/2$. Inset in (a) is a comparison of corrected demagnetization factor as a function of $R_1$ used in the DVPM computed by interpolation (solid red) and from demagnetization energies (black).  (c) The computed initial susceptibility is compared against simulation for disks varying radius (R) and thickness (L). Squares denote $R=250\thinspace$nm, circles $R=500\thinspace$nm, and triangles $R=1800\thinspace$nm. All simulations use $M_S=$ 800$\thinspace$kA/m except the $R=1.8\thinspace\mu$m which use $M_S=$715$\thinspace$kA/m. (d) Using the same simulation parameters the frequency of the gyrotrpic mode was computed. The legend in (c) applies to panel (d) as well. Only the DVPM agrees well with both the initial susceptibility and the gyrotropic frequency.  For (c) and (d) calculations were performed holding $R=$500$\thinspace$nm with variable thickness. All DVPM calculations in panels a-d use $R_1=$R/2. }

\end{figure*}

The TVM with a tangential boundary condition neglects the translation mode of displacement. There is, however, an overestimate of deformation that compensates, permitting good estimates of low field core displacement. However, the lack of translation introduces  a pervasive underestimate of the magnetic moment that develops near the edge of the disk, leading to an underestimate of magnetic susceptibility (Figure \ref{fig2} a). 

\subsection{Construction}

\begin{figure*}[H]
\includegraphics[scale=1.0]{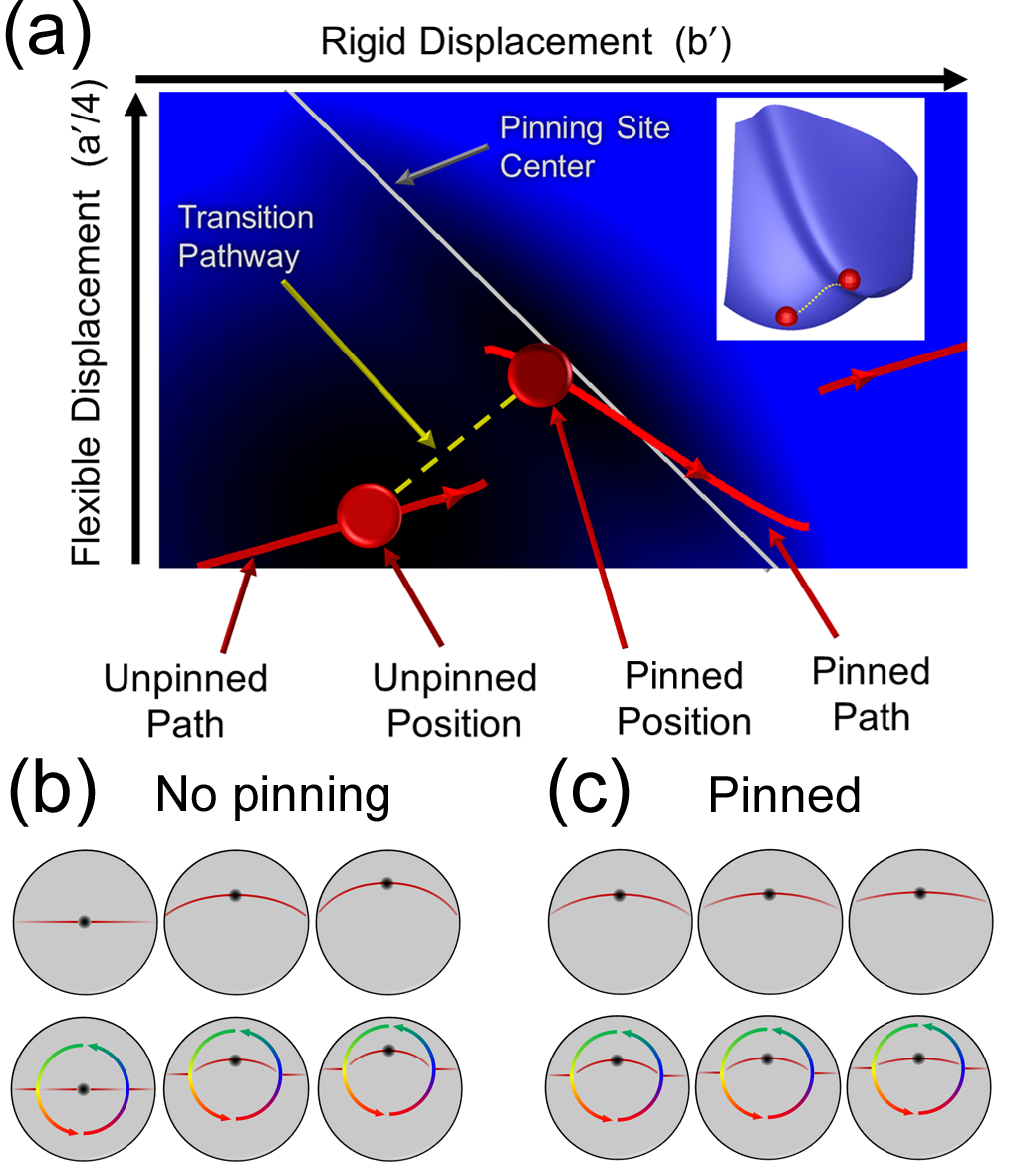}
\caption{\label{fig2new} (Color available in online version.) (a) A 2D plot of the energy landscape (color gradient) in the DVPM computed for a $1\thinspace\mu$m diameter, $40\thinspace$nm disk is shown as a function of both rigid displacement of the RVM annulus ($b'$) and flexible displacement of the central TVM region ($a'/4$ for $R_1=R/2$) for a fixed field value of $h=0.025$ using an $M_S=800\thinspace$kA/m. This is the 2D landscape that must be solved to find the lowest energy combination of $b'$ and $a'/4$ for a particular vortex displacement. The energy landscape chosen features a single pinning site at $b'+a'/4=0.21$. At $h=0.025$, two local minima exist, one inside the pinning site, one outside. As field is changed, the positions of these minima shift along the paths displayed above, allowing the DVPM to describe flexing and rigid displacement in unpinned and pinned situations. Inset at top right is a 3D representation of the potential at $h=0.025$, the two local minima and the transition pathway. (b, c) The schematics depict the field evolution of the deformation in magnetization distribution using the $|M_y|=1$ contour of a simulated disk (at top) and the DVPM (at bottom). (b) shows the no pinning case and (c) a case with strong core pinning. Note that in both the simulated and DVPM case, deformation decreases when the core is pinned, and increases when it is unpinned. This matches the computed pathway evolution shown in panel (a).}

\end{figure*}

With the two contributing models introduced, the DVPM  may now be constructed by dividing the disk into two regions, an outer annulus described by the RVM, and an inner disk of radius $R_1$ described by the TVM with no side charges.  The RVM annulus provides a representation of the outer region magnetization with the capability of translating rigidly independent of the core position. The TVM central region shifts with the RVM outer annulus but provides a flexible region to permit the core to advance or to lag on account of pinning, while still directly coupling the in plane magnetization to the core. In this construction, the RVM provides the increased magnetic moment from the outer sections of the disk, while the TVM core allows computation of the energy stored in the dipole-exchange spring and its effects on the core. In the absence of pinning the RVM annulus translates and the TVM center deforms, recreating the physical situation apparent in simulation (Figure \ref{fig2new} b). When the core is in a pinning site, the rigid annulus continues to deflect, which leads to a reduction of the flexing inside the TVM core in order to maintain the approximate vortex position (Figure \ref{fig2new} c). Again, the physical behavior is qualitatively reflected in the model. 

The use of the RVM for the outer region is motivated by the observation that a significant component of the magnetization distribution evolution can be described by translation. Additionally, the susceptibility correction means that the RVM is the best model for simultaneously predicting the net magnetization of the disk and the core position. Inside, the tangential boundary condition of the TVM maintains a piecewise continuity of the magnetization distribution \footnote{Inverting the TVM and RVM, so that the TVM is outside the RVM will result in the poor magnetization performance of the TVM dominating the model, an inability to restore circular symmetry and no natural boundary condition that can piecewise match the magnetization distributions between the two components.}. 

\subsection{Solving the DVPM}

The clear challenge in developing this piecewise model, is modifying the coupling between the two models such that a reduced central flexible region correctly describes the energetics of flexing over the entire disk. The correct computation of the total energetic capacity of the dipole-exchange spring  for a given core displacement will be critical to the model's ability to compute the influence of the spring and any consequent changes during pinning. It is also important to correctly compute the energetic cost of translation. Effectively, the optimal coupling of the two models will compute the correct proportion of translational displacement, and deformation based displacement while also computing the correct energetic cost of the deformation. 

As previously noted, the two contributing models represent the limiting cases of translation only, and of no translation. The RVM uses the susceptibility correction to take into account the displacement due to deformation, while the TVM overestimates displacement due to deformation, compensating for the lack of translation. A combined model, with no changes to either component results in artificially increased predicted values of both the positional and magnetic susceptibilities of the vortex state. Therefore, the energetic cost of displacement (and consequent increase in magnetic moment) must be increased for both of the components to compensate. The reduced size of the TVM region already represents a reduced flexing potential. Exacerbating this makes little sense. Therefore the energetic cost of the rigid annulus should be also increased.

The magnitude of this correction for the outer annulus may be computed by considering the demagnetization energy, which provides the dominant energy contributions in each model. Inside the TVM region, the vortex is displaced in an energetic well, dominated by the volume demagnetization charges. The spring force governing the core position in this well is approximately given by the coefficient of the derivative of the energy of the induced volumetric demagnetization charges with respect to the displacement parameter $a/2$. The ability to move the vortex via the RVM shell must be reduced by an equivalent amount by increasing the stiffness of the well constraining rigid translation. This may be incorporated by an equivalent increase in the demagnetization factor of the outer shell. The equivalent change of the RVM shell demagnetization factor necessary to rebalance the energetic cost of vortex displacement is $(R_1/R)^2 (2R_1/L)F_1(R_1,L)$.

The physical accuracy of this coupling approach may be grounded by considering the fundamental meaning of the susceptibility correction incorporated into the RVM. The susceptibility correction implicitly couples the two sections through the demagnetization energy. In the RVM, the susceptibility correction uses material susceptibility to estimate the net decrease of the demagnetization energy due to the reduction of side charges, and the introduction of volume charges parameterizing the real world magnetization flexing. But the TVM is already flexible, and therefore including the susceptibility correction for the shell, in addition to a central TVM region reduces the demagnetization cost of translating the magnetization distribution too much. Ideally, the central region would parameterize only the gentle bending of the contours that leads to the deformation displacement, while the steeper edge bending that reduces translation cost would still be parameterized by the susceptibility correction. This implies that the susceptibility correction should be decreased in magnitude proportionally to the size of the TVM region. Comparing a linear interpolation between the corrected ($F_c$) and uncorrected ($F_{nc}$) demagnetization factors of the form $F(L,R,R_1) = (1-R_1/R)F_c (L,R)+(R_1/R)F_{nc}(L,R)$ to the computed change in the RVM shell demagnetization accuracy above shows  that the deviation between the two does not exceed 10\% over the range $R_1/R = $0 to 1\footnote{Here interpolation is chosen over the estimated rebalancing energy to maintain the limiting values at the susceptibility-corrected demagnetization factor and the uncorrected value.}. This corroborates the application of susceptibility correction and its physical interpretation in the context of a rigid model. 

Using the interpolated demagnetization factor, the energy of the combined piecewise model may be written,

\begin{equation}
\frac{E_{tot}}{\mu_0M_s^2V}= \frac{\beta'}{2}b'^2-h (b'-\frac{b'^3}{8}) +\gamma(\frac{\alpha'}{2}a'^2-\xi ha'),
\label{eq:Edvpm}
\end{equation}

where $\beta' = F(L,R,R_1) - R_o^2/R^2$ and $\alpha' = R_1 F_1(L,R_1)/L-R_o^2/2R_1^2$.  Here $b'$ is the normalized displacement of the outer RVM shell, and $a'/2 =\Delta r_1/R_1$ is the central TVM core displacement normalized to $R_1$. The factor $\gamma = R_1^2/R^2$ scales the energy contributions accordingly. All other symbols remain as before. The total core displacement is $s= \Delta r/R + \Delta r_1/R = b' + R_1a'/2R$ with the same expressions for $b'_o(h)$ and $a'_o(h)$ as before, but with $\beta'$ and $\alpha'$ replacing $\beta$ and $\alpha$ respectively. The corresponding magnetization is $m_o(h) = b'_o(h) -b'_o(h)^{3}/8 +\gamma \xi a'_o(h)$. For small displacements, the third order terms for the RVM shell may be dropped to make a simplified version of the model.

Only one free parameter remains, the radius of the inner TVM section. The influence of $R_1$ is mitigated by the coupling of the models using the susceptibility correction, however the choice of $R_1$ is not entirely arbitrary. The susceptibility correction interpolation maintains the energetic cost of total vortex displacement, but $R_1$ still determines how much of the deformation based displacement is explicitly accounted for by the inner region and how much is attributed to the shell, via the remaining correction. The constructed model can behave as the RVM in one limit ($R_1=0$), or the TVM in the other limit ($R_1=R$). Consequently, the model can exhibit the failings (and successes) of the RVM in one limit, and the TVM in the other. Optimal computation of the properties of the vortex state require an intermediate $R_1$. This can be estimated by minimizing the deviation of the DVPM from the successful predictions of the RVM ($m(h)$, $s(h)$) and the TVM (gyrotropic frequency). In general, a reasonable agreement with all three parameters can be found for $R_1$ values of approximately $R/2$. In more detail, the optimal $R_1$ has a weak dependence on the radius of the disk. Semi-empirically, the optimal $R_1 = R(0.6-(5/3)(L/R))$ was computed for disks of varying radius and $40\thinspace$nm thickness. 

\section{Application to the Ideal Disk}

 From Equations \ref{eq:Ervm}, \ref{eq:Etvm}, and \ref{eq:Edvpm} the ideal disk behaviour of each model may be computed and compared to Landau-Lifshitz-Gilbert micromagnetic simulation\footnote{All simulations were performed using version 2.56d of the LLG Micromagnetics software package http://llgmicro.home.mindspring.com/.}. To mimic quasistatic behaviour, time integration with a damping factor of 1.0 was used. All simulations were performed on a 2-D 5$\thinspace$nm$\times$5$\thinspace$nm$\times$thickness grid using an exchange stiffness constant of $1.05\times 10^{-11}$J/m, with $M_S$ values between 700kA/m and 800kA/m and either 20$\thinspace$nm or 40$\thinspace$nm thickness.  All calculations with the model used an exchange length of 5.85$\thinspace$nm and $M_S$ values matching the simulations. 

Comparison of the $m(h) = M(H/M_S)/(\mu_o M_S V)$ and $s(h)= \Delta r(H/M_S)/R$ curves are shown in Figures \ref{fig2} a and b. Clearly the susceptibility-corrected 3rd order RVM provides the best estimate of both magnetization and vortex displacement as a function of field, while the uncorrected version exhibits the poorest performance.  Both the DVPM and TVM provide good estimates of vortex position with field for displacements up to $R/2$, but only the DVPM simultaneously gives a good description of the magnetization. 

Two other metrics have been applied to evaluate the performance of the analytical models near zero field in past work: initial susceptibility, and the frequency of the lowest order excitation mode of gyrotropic vortex motion.  Both of these parameters primarily depend on the aspect ratio of the disks. Initial susceptibility is easily calculated from $m_o(h)$ for each model.  Using the collective coordinate approach \cite{Thiele}, it may be shown that the gyrotropic mode frequency is $f_o=\kappa/2\pi G$ where $G=2\pi L M_S/\gamma_o$ with $\gamma_o=1.76\times10^{11} s^{-1}T^{-1}$, and $ dE_{tot}/d r = \kappa r$ \cite{TVMJAP}.   For the RVM and TVM, $\kappa$ is $\beta$ and $4\alpha$ respectively. For the piecewise combined model, $\kappa$ may be computed in the unpinned and zero field case

\begin{equation}
\kappa= \frac{4(\alpha+\gamma \xi^2 \beta)\beta \alpha}{(2\alpha +\rho \xi \beta)^2}.
\label{eq:kappa}
\end{equation}

Comparisons between simulation and computed results for initial susceptibility and gyrotropic mode frequency are shown in Figure \ref{fig2} c and d. Dynamic simulations were performed using a realistic damping factor (0.02) but otherwise matched the parameters used in the previous simulations. The poor performances for magnetization description of the TVM and uncorrected RVM manifest as incorrect estimations of the initial susceptibility. However, both approach the simulation results for squat disks, corroborating previous results\cite{GusPRB, TVMJAP} and demonstrating the general utility of these models. By comparison, the DVPM and corrected RVM provide excellent estimates of initial susceptibility for all aspect ratios investigated.  Previously, only the TVM has provided reasonable estimates for the gyrotropic frequency of the vortex state while the RVM has provided poor estimates. The success of the TVM is reproduced here, as is the failure of the uncorrected RVM. The susceptibility correction improves the RVM prediction, however it fails to match the performance of the TVM in the prediction of $f_o$. However, the DVPM provides comparable performance to the TVM for low aspect ratios and improved performance with more squat ideal disks.

\section{Application to Pinning}

Having demonstrated the performance of the piecewise model in a perfect disk, pinning may now be considered. Adding pinning to the models is accomplished by adding functions of the form $E_p(b’ + R_1 a’/2R-X_p)$  for a pinning site located at $X_p$ to equation \ref{eq:Edvpm}, or of the form $E_p(b - Xp)$ for the RVM in equation \ref{eq:Ervm}. For the RVM case, simply solving for the minima in energy permits a full solution of the problem. For the DVPM, the 2-D optimization required makes the problem more complicated, however this is critical to the success of the model.  Plotting the pinning energy, pinning sites appear as linear troughs in $b'$-$R_1 a'/(2 R)$ space (Figure \ref{fig2new} a). This permits a simplification of the optimization process by the consideration of pinning site coordinates defined by $b’ = i\thinspace sin (\theta) + j\thinspace cos(\theta) $ and $a’ = i \thinspace cos(\theta) - j\thinspace sin(\theta)$ for $\theta = tan^{-1}(2R/R_1)$. Switching to $i$ and $j$ coordinates allows independent minimization and simplifies the problem.   The position and existence of local minima inside and outside of pinning sites evolves with changing applied field (Figure \ref{fig2new} and Supplementary Movie M3). Sometimes bistable states exist, and when they do so, there is inevitably a transition pathway between the two extant minima that passes over a saddle point. Applying a 2D optimization repeatedly while changing the field permits computation of the values of $b’$ and $a’$ for all minima and saddle points. This in turn permits computation of the quasistatic pinned  and unpinned magnetization and vortex position, while locating the saddle points separating minima allows computation of the energy barriers separating bistable states.

\begin{figure*}[H]
\includegraphics[scale=1.0]{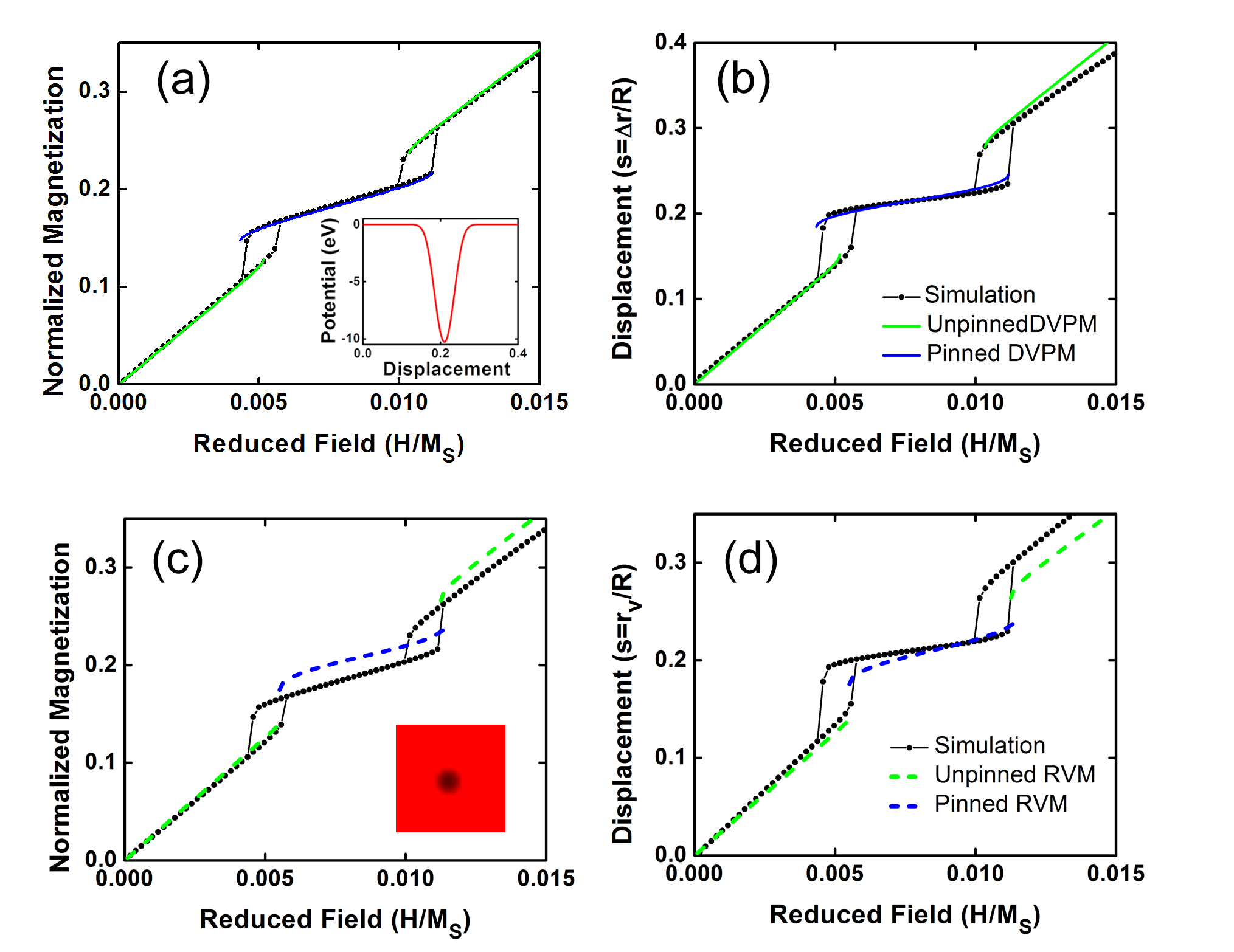}
\caption{\label{fig3} (Color available in online version.) The m-h (a) and s-h (b) curves for a field sweep up and down are compared against simulation for a 1$\mu\thinspace$m diameter, 40$\thinspace$nm thick disk with a single pinning site located at 105$\thinspace$nm with an energy profile estimated from the simulation (inset in (a)). For comparison, the same pinning site is added to the susceptibility corrected, one parameter RVM. The computed m-h (c) and s-h (d) curves provide much worse agreement including a larger (($>10\thinspace$nm) positional error, and more importantly, a complete absence of hysteresis on entrance and exit from the pinning site.  In the simulation the pinning site is included as a region of suppressed $M_S$ (inset in panel (c)) with diameter 40$\thinspace$nm, where at the center $M_S=$550$\thinspace$kA/m.}

\end{figure*}

The critical feature of this minimization process is that the coordinate $a$ may evolve non-monotonically with increasing field (Supplementary Material Movie M3) matching the qualitative non-monotonic evolution of the flexing of the magnetization distribution visible in simulation (Supplementary Movies M1 and M2).  

Micromagnetic simulations were used evaluate the pinning performance of the DVPM and, for comparison, the corrected RVM. The same simulation parameters were used as in the previous simulations. Pinning sites are mimicked using approximately circular regions of depressed saturation magnetization to modify the energy landscape of the disk (Figure \ref{fig3} c inset). This leads to two contributions to pinning energy, the reduced exchange energy of the core in the low $M_S$ region, as well as reduced demagnetization energy when the core is centered on the site. The energetic profile of the pinning site can be approximated by considering the convolution of a 2-D Gaussian at various offsets with the profile of the $M_S$ variation (Fig \ref{fig3}a, inset). The Gaussian effectively approximates the exchange energy density of the core, as well as the $M_z$ profile, providing an estimate of how the two energy contributions change as the core shifts relative to the pinning site. Here a full width half max of 17.2$\thinspace$nm is used for the Gaussian approximation.  

\begin{figure*}[H]
\includegraphics[scale=1.0]{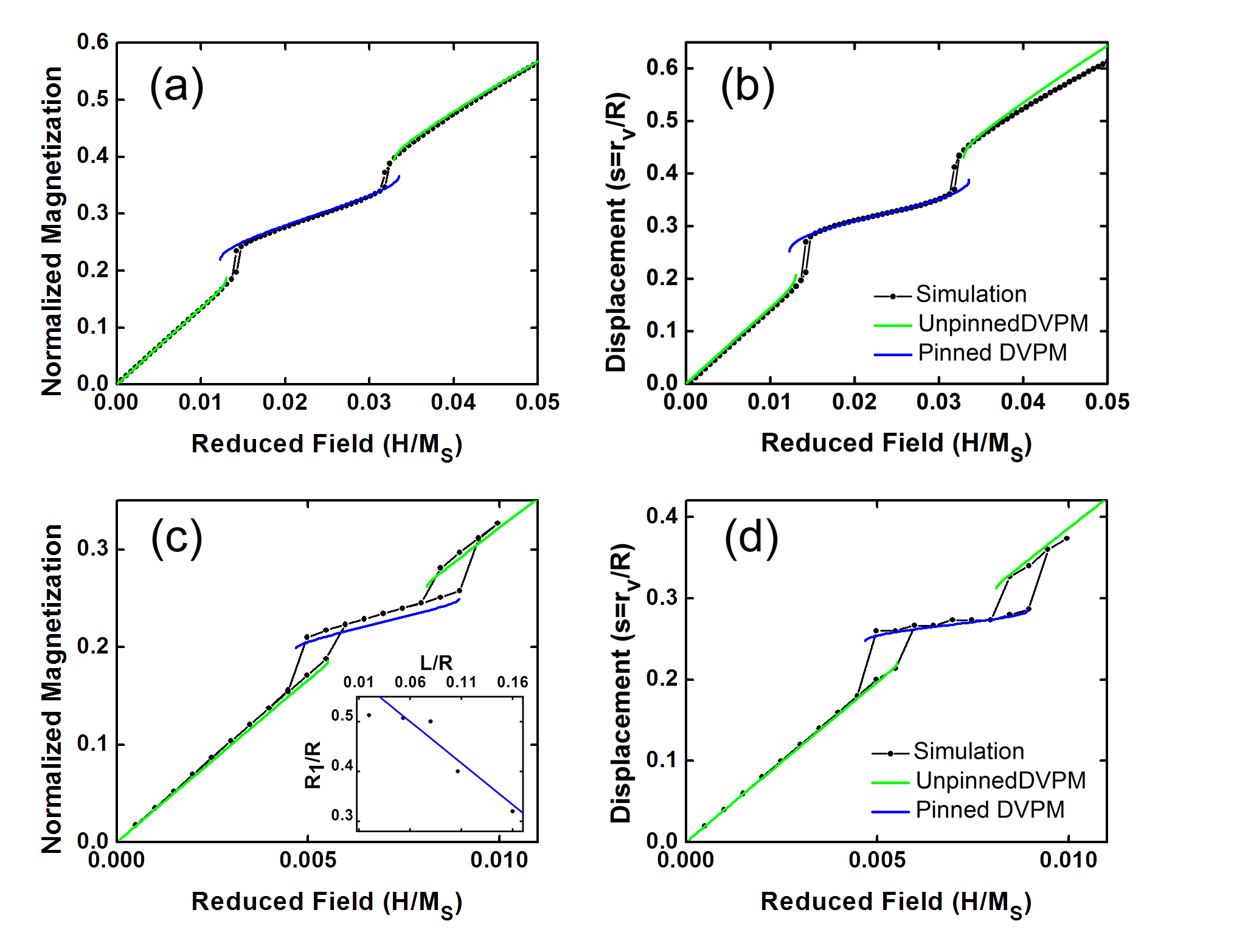}
\caption{\label{fig4} (Color available in online version.) (a) The DVPM m-h result is compared to a simulation of a 500$\thinspace$nm diameter, 40$\thinspace$nm thick disk with $M_S =$700$\thinspace$kA/m and a single pinning site located 80$\thinspace$nm away from the center. (b) The computed normalized displacements are compared for the same simulation. For this comparison with $R=250\thinspace$nm, $R_1 =$80$\thinspace$nm provides the best estimate of hysteresis loop width. The inset shows a plot of the optimal $R_1$ value found by comparison to simulation as a function of aspect ratio (black points). The blue line is the optimal $R_1$ value computed by minimizing deviation of the DVPM from the RVM and TVM for initial susceptibility, $ds/dh$ and gyrotropic mode computed for $40\thinspace$nm thick disks. (c) Comparison to a simulated 1500$\thinspace$nm diameter, 40$\thinspace$nm thick disk with a single pinning site at 200$\thinspace$nm from center shows that the DVPM begins to underestimate the magnetization as disk size increases. (d) The computation of the vortex position and hysteresis loop width remains accurate. For $R=750\thinspace$nm, an $R_1$ value of 375$\thinspace$nm was used. }

\end{figure*}

The performance can be evaluated by three metrics: the pinning site position error, the width of the minor hysteresis loops associated with pinning and depinning, and the combined computed pinned differential magnetic and positional susceptibilities.  Figure \ref{fig4} a-d show results for a 1 micron diameter, 40$\thinspace$nm thick, disk compared to the DVPM and the 3rd order RVM. The DVPM accurately captures both differential susceptibilities while the RVM fails to capture the positional slope. Both models feature effective position shifts of the pinning site. The DVPM agrees best with the simulation for a pinning site shifted 2.5$\thinspace$nm further from center than the actual simulation (107.5$\thinspace$nm instead of 105$\thinspace$nm), while a shift greater than 10$\thinspace$nm  is best for the RVM (at 115$\thinspace$nm instead of 105$\thinspace$nm).  Most importantly however, the computed entrance and exit hysteresis loops agree closely for the DVPM, but are almost non-existent for the RVM. The deformation allowed by the DVPM permits the vortex to move ahead into the site, and linger in the site at a lower energy cost than the rigid model. 

Disk sizes between 500$\thinspace$nm diameter/40$\thinspace$nm thick and 2000$\thinspace$nm diameter/20$\thinspace$nm thick were simulated with identical $M_S$ variation pinning sites (Fig \ref{fig4}). The value of $R_1$ used makes a significant difference in computing pinning effects in comparison to unpinned behavior. Changing $R_1$ has a weak influence on the computed m-h and r-h curves, mitigated by the coupling approach used.  More importantly, the value $R_1$ dictates the energetic cost of displacing the core via the exchange-demagnetization spring. Reducing the proportional value of $R_1$ stiffens the spring. This in turn has a significant effect on pinning and depinning barriers. In the previous section, errors between the DVPM and its component models were analyzed to determine the optimal $R_1$, providing an $R_1$ estimate independent of simulations. In the case of pinning, however, no model is adequate for comparison, and consequently it is best to determine an optimal $R_1$ value by comparison to well-defined pinning simulations. As in the comparison made to the RVM and TVM, in general the value $R_1=R/2$ provides reasonable results. However, for disks below 1$\mu$m in diameter (for 40$\thinspace$nm thickness), reduced $R_1$ values provide better pinning performance, reflecting the increasing rigidity of smaller disks. A comparison between the optimal $R_1$ values computed by error minimization against the RVM and TVM, and also by comparison with simulation, is shown in the inset of Figure \ref{fig3}d. Each optimization method returns the same qualitative trend of decreasing $R_1/R$ with the value of $R$. 

The DVPM was found to give good estimates of hysteresis width and vortex position for all disk sizes when an optimized value of $R_1$ was used. For disks significantly larger than 1$\thinspace\mu$m in diameter, at 40$\thinspace$nm thickness, the pinned magnetic differential susceptibility was found to be underestimated. Figure \ref{fig4} shows a 500$\thinspace$nm and 1500$\thinspace$nm diameter result for comparison.  The $R_1$ from error minimization provides a reasonable estimate for situations where simulation is not possible, however for the most accurate computation of pinning effects, constructing a known simulation is preferable. For large and thin disks, $R_1=R/2$ provides better agreement with simulation. This reflects the fact that in low aspect ratio disks that are very large compared to the exchange length, the character of flexing in the magnetization distribution will change to include more complex deformations beyond the scope of the TVM approximation used. It should be noted that for a given disk aspect ratio, once the optimal $R_1$ is computed from a single simulation with well known pinning parameters, the $R_1$ value is then fixed. This permits computation of the effects of arbitrary pinning potentials or even fitting magnetization curves to extract information about the pinning potentials. 

\subsection{Two Dimensional Pinning Potentials}

\begin{figure*}[H]
\includegraphics[scale=1.0]{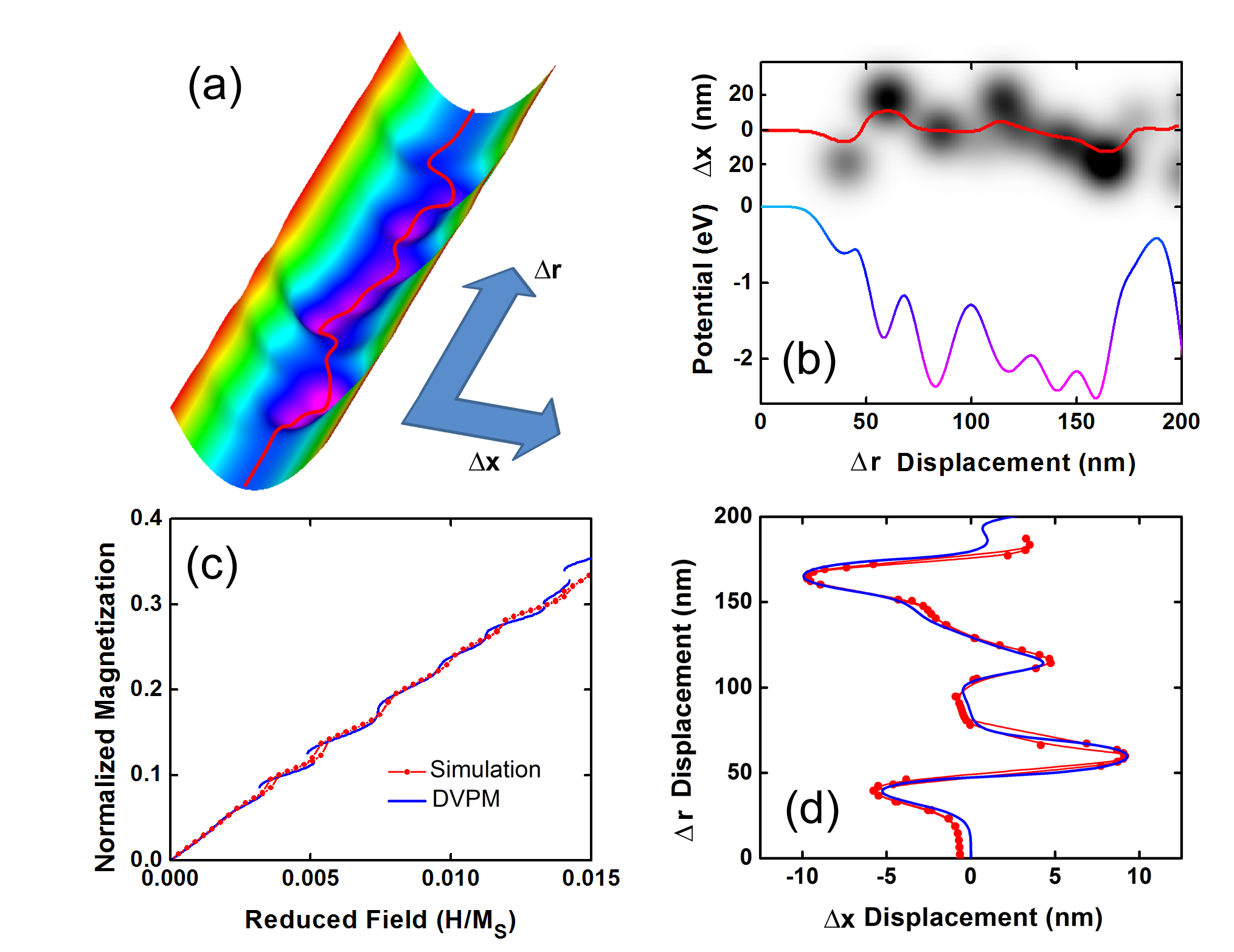}
\caption{\label{fig5} (Color available in online version.) (a) A 3-D plot shows the pinning site potential  combined with the harmonic potential for deviations orthogonal to the path defined by the applied field. The computed path of minimum energy, $\Delta x(\Delta r)$ is plotted as a red thread. (b) At top the 2-D pinning potential is plotted with the computed minimum energy path (red line). Below, the equivalent 1-D potential is presented. The equivalent potential incorporates contributions from both the pinning potential and harmonic trough. The color gradient on the potential line matches the color scale in panel (a). (c) The magnetization curve computed from the potential in (b) is compared against a simulated curve incorporating $M_S$ suppressed regions with the same 2-D distribution. The depth of the 2-D potential used in the model calculations is estimated from the simulation. Agreement is very close, though some deviations show up as the vortex displacement increases and a large energy change is encountered. (d) The simulated $\Delta x(\Delta r)$ is compared to the computed minimum path showing excellent agreement.}

\end{figure*}

The DVPM provides excellent performance in the description of ideal disk behavior and pinning for idealized simulations. However, in application to real samples, the treatment of pinning sites located directly along the pathway followed by the vortex in the absence of pinning as it is deflected by field is limiting. As noted in recent numerical simulation work on pinning\cite{Wysin}, a more realistic case is to consider pinning sites near, but not centered on, the field-defined path. This can be incorporated into the 1-D model presented here by computing the 1-D equivalent potential of the actual 2-D path followed by the vortex. Deviations orthogonal to the path defined by the applied field have an energy cost approximated by $\kappa \Delta x^2$ where the value of $\kappa$ is given by equation \ref{eq:kappa}.  Since the magnetization induced by these deviations is orthogonal to the applied field, the energy is effectively static and can be summed with a 2-D distribution of pinning sites to form a trough guiding the vortex through the 2-D energy landscape. It is then possible to compute the minimum energy pathway $\Delta x_o(\Delta r)$ that the vortex will follow as it is deflected (Figure \ref{fig5}a). Computing the total static energy, pinning plus the trough energy, $E(\Delta x_o )$ yields an equivalent 1-D potential as a function of $\Delta r$ (Figure \ref{fig5}b).  This potential can then be summed, as the Gaussian pinning sites were previously, with the potential for a perfect disk including field. Solving for minima as before allows computation of the evolution of the magnetization and vortex position in the 2-D potential. 

This approach is applied to a simulation that incorporates a 2-D distribution of 10$\thinspace$nm diameter pinning sites with various values of suppressed $M_S$ near the field-defined path. As before, the pinning sites are incorporated into the model as Gaussian wells with depths estimated from the simulation and profiles computed by convolving a Gaussian with the profile of the $M_S$ variation. The computed 2-D path agrees well with the vortex position extracted from simulation, as does the computed magnetization (Figure \ref{fig2} c and d). Some disagreement is noted as the deflection increases close to the effective $R/2$ limit of the model, and the vortex passes over a large barrier.  In this computation, a sparse 2-D distribution of sites ensures a unique $\Delta x_o (\Delta r)$, however, in principle this approach can be extended to bistable states in $\Delta x$ by consideration of multiple vortex tracks. Computation of the energy barriers separating tracks, however, would require a more complete minimization.

\section{Conclusion}

The piecewise approach applied to develop the DVPM yields a highly functional analytic model that makes quantitatively accurate predictions of a wide variety of properties of a vortex in a disk.  Most notably, it provides a powerful description of vortex core pinning and provides greater physical insight into the behavior of the vortex during pinning. The model holds promise as a tool in probing the modification of pinning in technologically pertinent thin films to better understand effects such as ion damage, while the piecewise approach demonstrated may, in future, be generalized to other geometries, permitting quantitative computation of device behavior without cumbersome simulation.

\begin{acknowledgments}
We are grateful for support from the Natural Science and Engineering Council of Canada, the Informatics Circle of Research Excellence,  the National Institute for Nanotechnology, the Canada Research Chairs program, Alberta Innovates, and the Canadian Institute for Advanced Research. 
\end{acknowledgments}

%

\section{Supplementary Material: A Deformable Model for Magnetic Vortex Pinning}

The first movie (M1) shows the magnetization distribution of the disk from the simulation that generated the data used in Figure 3 in the main text . The magnetization direction is indicated by the color. Red indicates magnetization in the positive $x$ direction, green is negative $x$, blue is positive $y$ and yellow is negative $y$. 

 \vspace{10 mm}

As the vortex is displaced by the field, the magnetization distribution warps away from the circularly symmetric initial vortex state. As the vortex interacts with the pinning site, the warping changes, and the flexing of the distribution decreases. This results from the continued biasing of the magnetization distribution outside of the core effectively mimicking continued displacement of the outer region while the core has reduced mobility. This can be challenging to see on the color scale. A second movie (M2) of the same simulation shows only the $y$ component of the magnetization in a contour plot. The non-monotonic evolution of the contours stands out more clearly. 

 \vspace{10 mm}

The third movie (M3) shows the evolution of the energy (color scale) of the vortex for a given location dependent on the parameters $a'$ and $b'$ computed in the model as magnetic field increases. The vertical axis is $a'$ and $b'$ is the horizontal coordinate. The single pinning site shows up as a trough across $a'$,$b'$ space. As the field increases, the unpinned vortex position (blue dot) shows increasing translation ($b'$) and deformation ($a'$), and a second minimum appears in the pinning trough (red dot). As the field increases further, the vortex position outside the pinning site continues to increase in both $a'$ and $b'$, but the pinned minimum decreases in $a'$. This effectively captures the behavior in the simulated movies M1 and M2. The flexing decreases, while the outer region continues to displace. 

 \vspace{10 mm}

In addition to providing the opportunity to determine the positions of energetic minima to compute the unpinned and pinned magnetization and vortex displacement curve as a function of field, the saddle point separating the two minima may also be computed, allowing calculation of the energetic barrier separating the two.

\end{document}